\documentclass[twocolumn,showpacs,preprintnumbers,superscriptaddress,amsmath,amssymb,prl]{revtex4}

\usepackage{graphicx}
\usepackage{dcolumn}
\usepackage{bm}

\begin{document}


\title{Entanglement assisted metrology}

\author{P. Cappellaro} \affiliation{Department of Nuclear Engineering, Massachusetts Institute of Technology, Cambridge, MA 02139, USA} 
\author{J. Emerson} \affiliation{Perimeter Institute of Theoretical Physics, Waterloo, ON N2J 2W9, Canada} 
\author{N. Boulant} \affiliation{Department of Nuclear Engineering, Massachusetts Institute of Technology, Cambridge, MA 02139, USA} 
\author{C. Ramanathan} \affiliation{Department of Nuclear Engineering, Massachusetts Institute of Technology, Cambridge, MA 02139, USA} 
\author{S. Lloyd} \affiliation{Department of Mechanical Engineering, Massachusetts Institute of Technology, Cambridge, MA 02139, USA} 
\author{D.G. Cory} \thanks{Author to whom correspondence should be addressed} \affiliation{Department of Nuclear Engineering, Massachusetts Institute of Technology, Cambridge, MA 02139, USA}


\date{\today}

\begin{abstract}
We propose a new approach to the measurement of a single spin state, based on nuclear magnetic resonance (NMR) techniques and inspired by the coherent control over many-body systems envisaged by Quantum Information Processing (QIP). 
A single target spin is coupled via the natural magnetic dipolar interaction to a large ensemble of spins. Applying external radio frequency (rf) pulses, we can control the evolution of the system so that the spin ensemble reaches one of two orthogonal states whose collective properties differ depending on the state of the target spin and are easily measured. We first describe this measurement process using QIP gates; then we show how equivalent schemes can be defined in terms of the Hamiltonian of the spin system and thus implemented under conditions of real control, using well established NMR techniques. We demonstrate this method with a proof of principle experiment in ensemble liquid state NMR and simulations for small spin systems.
\end{abstract}


\maketitle

The measurement of a single nuclear spin state is an experimentally challenging task, that once solved could yield useful applications as well as valuable physical insights.
Potential applications include nuclear spintronics devices \cite{Spintronics}, biomolecular microscopy 
and spin based QIP \cite{Kane}. 
Many methods have been proposed to increase the sensitivity of NMR:
force detection \cite{MRFM} and near field optics \cite{wrachtrup}, for example, have reached the limit of single spin detection, but the measurement of the quantum state of a single nuclear spin has not yet been achieved. 

We present here a new approach that relies on coherent collective properties of a quantum system. 
We propose to use the interaction of the spin of interest (target spin) with an ensemble of spins to amplify its signal up to the point where it is detectable by the usual inductive means. This method requires the ability to control about $10^6$ spins (the current low temperature detection limit for NMR) by inducing a coherent dynamics, modulated by the interaction with the target spin. Because of the challenges in creating a macroscopic entangled state, in the short term we expect our method to be used to enhance sensitivity, with a system of about hundred entangled spins (already available experimentally). As no physical limitations prohibit to enlarge the system to the size required for single spin state measurement, this remains however our ultimate goal.

The ensemble of spins, forming a device that we call Spin Amplifier, is first prepared in a highly polarized state; it is then put in contact with the target spin, which has already collapsed into one of the states $|0\rangle$ or $|1\rangle$ 
\cite{endnote1}.

To illustrate how a collective measurement can provide knowledge of the single spin state, consider a simple quantum circuit, consisting of a train of Controlled Not (\textsc{c-not}) gates between the target spin and each of the Amplifier spins. Since the \textsc{c-not}s act on the fiducial state $|00\dots0\rangle$, this amounts to effectively copying the target spin state onto the Amplifier spins \cite{endnote2}. 
At the end of the circuit, the measurement of the Amplifier magnetization along the z-direction ($M_z \propto \langle \Psi_A|\sum_i^n\sigma_z^i|\Psi_A\rangle$) will indicate the state of the target spin. 

Although the measurement involves only collective properties of the Amplifier, this first scheme demands that the ensemble spins are independently addressable and that they all interact with the target spin. 
\begin{figure}[b]
\centering
	\includegraphics[scale=0.8]{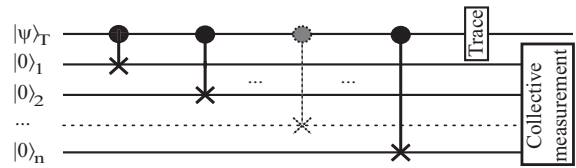}
 \caption{\textit{Scheme 1}: Series of \textsc{c-not} gates between the target and Amplifier spins. A collective measurement is sufficient to detect the state of the target spin.}
\end{figure}

Instead of imposing these requirements, we can develop equivalent schemes that connect better to the available control. Relying only on entanglement among the Amplifier spins and on the evolution given by the internal Hamiltonian and collective rotations, these schemes can be realizable in the near term. 
Entanglement among particles in an ensemble has been shown to produce changes in macroscopic observables \cite{Khitrin} and to enhance the signal-to-noise ratio in spectroscopy \cite{wineland} up to the Heisenberg limit: The maximally entangled state can acquire the phase information with optimal sensitivity, since an $n$-qubit cat-state evolves $n$ times faster under a collective evolution. In a similar way, we want to create an entangled state that is the most sensitive to the action of the target spin.
In scheme 1, where the Amplifier spins remain in a factorable state, the interaction with the target spin produces only local changes on individual spin states. 
\begin{figure}[t]

	\label{Entanglement1}
	\centering
			\includegraphics[scale=0.75]{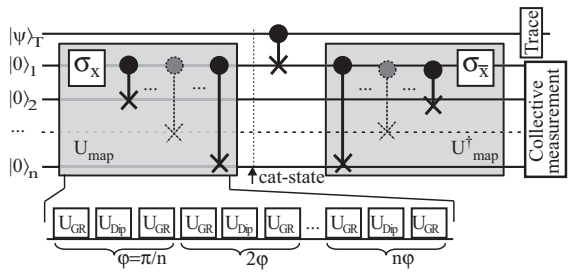}
	 \caption{Entanglement permits to use only one local action of the target spin. The entangling operator $U_{map}$ can be created with gates on single spins (above) or with a phase modulated sequence, using only the internal dipolar Hamiltonian and collective rf pulses to create the Grade Raising operator (below).}
\end{figure} 
On the other hand, with the creation of a macroscopic entangled state \cite{Leggett}, the Amplifier
is globally affected by a single interaction with the target spin: The propagator creating the cat-state performs an effective change of basis to a reference frame where the local \textsc{c-not} gate is a global operator on the Amplifier. 
We illustrate the role of the entanglement with a second scheme (Fig. 2).
Here the fully polarized state $|00\ldots0\rangle$ is first transformed into the cat-state $\frac{1}{\sqrt{2}}(|00\ldots0\rangle-i|11\dots1\rangle)$ by a $\pi/2$ rotation about $\sigma_x$ and a series of \textsc{c-not}s; then, we invert the state of the first Amplifier spin, conditionally on the state of the target spin.
When we next invert the evolution to undo the entanglement, the \textsc{c-not} and $\sigma_x$ gates bring back the Amplifier to the initial state if the target spin is in the $|0\rangle$ state. Otherwise we obtain the state: $|1\rangle_T|111\ldots 1\rangle$. 
As in the previous scheme, measuring the Amplifier magnetization provides information about the target spin state. Notice that we have invoked only one interaction between the target spin and a privileged spin in the Amplifier, a more practical requirement, given the locality of any spin-spin interaction. 

We have implemented this scheme on a small QIP NMR liquid system, where the target spin is represented experimentally by a macroscopic ensemble of spins. Although their state is detectable, it is measured only indirectly, following the scheme proposed. The single protons of a $^{13}C$ labeled Alanine molecule ensemble ($C^3H_3-C^1\textbf{H}^T(NH_2)-C^2OOH $) are the target spins, while the 3 carbons compose the Amplifier. The spins interact via the scalar coupling $\sum_{ij}J_{ij}\sigma^i\cdot\sigma^j$. 
\begin{figure}[hb]
	\label{Experiment}
	\centering
			\includegraphics[scale=0.5]{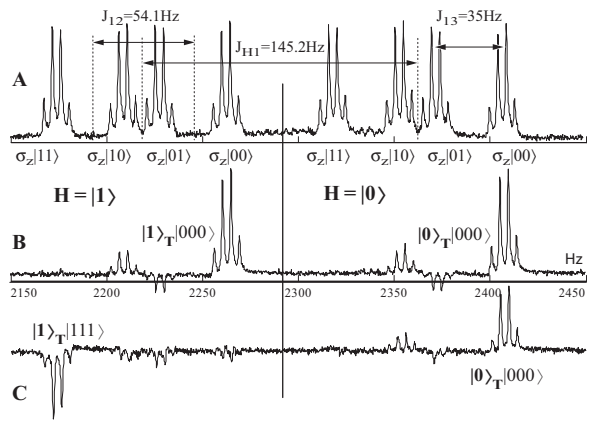}
	 \caption{\textit{Experimental Results:} \textbf{A}. Spectrum of the first carbon at thermal equilibrium, showing the coupling with the proton and the other 2 carbons (The methyl group produces the multiplet splitting). \textbf{B}. Initial State: Pseudo-pure state of the 3 carbons. \textbf{C}. Final state: The polarization of the carbons has been inverted (left) indicating a coupling to the target spin in state $|1\rangle$, while it is unchanged for spins coupled to target spin in the $|0\rangle$ state (right).}
\end{figure}
As it can be seen from the spectrum of the first carbon (Fig 3.A), the couplings with the proton are completely resolved, so that we can separate the signal arising from carbons coupled to protons in the $|1\rangle$ state (left) or in the $|0\rangle$ state (right). 
Before applying the circuit of Fig. 2, we put the proton spin into the identity state and prepare the carbons in the pseudo-pure state $|000\rangle$ \cite{forschritte}. The proton state is a mixture of the two states $|0\rangle\langle0|$ and $|1\rangle\langle1|$, thus we can effectively perform two experiments in parallel, and read out the outcomes from just one spectrum.  The gates in Fig. 2 (above) are implemented using the strongly modulated pulses developed in \cite{softpulses}. The experimental results show that the polarization of the three carbons is inverted conditionally on the state of the proton, giving an indirect measurement of its state.

Extending this scheme to larger systems is experimentally challenging, since we need to address individual spins. Yet we can relax some of the requirements on control, thus permitting different implementations, if we allow for more freedom in the final state of the amplifier.

The signature of a successful scheme is that it produces observable contrast, defined as: $C=\frac{M_z^0-M_z^1}{M_z(0)}$, $M_z^0$ and $M_z^1$ being respectively the magnetizations obtained when the target spin is in the state $|0\rangle$ or $|1\rangle$ and the initial magnetization is $M_z(0)$. 
The previous schemes provide the maximum contrast $C=2$. Even if their characteristics are quite different, they are equivalent since they create the same effective propagator: $E_T^+\prod_{i=1}^n \sigma^i_x+E_T^-$ (where $E^+=|1\rangle\langle1|$ and  $E^-=|0\rangle\langle0|$). This is the only operator (up to phase factors or constants of the motion) that results in the maximum contrast. 
If we accept a lower, but still observable, contrast ($C \approx 1$) the evolution we should implement, $U_{map}U_{cnot}U_{map}^\dag=E^+_T\mathcal{A}+E^-_T$, is such that the operator $\mathcal{A}$ instead of inverting the magnetization should just bring it to zero. In this case, $\mathcal{A}$ can have a much more general form than the one seen before ($\mathcal{A}=\prod_{i=1}^n \sigma_x$), as for example a superposition of $n/2$-quantum operators $\mathcal{A}=\sum_{\{i\}}(\prod_{i=1}^{n/2}\sigma_i^+ +h.c)$.

A realizable scheme, taking advantage of this flexibility, is a propagator that still creates a cat-state, but using only collective control and evolution of the Amplifier, such as it is available in a dipolar coupled spin system.

Techniques for generating entanglement have been developed in the context of spin counting experiments \cite{SpinCounting2}. The aim of these experiments is to calculate the size of a spin cluster by measuring its entanglement (or more precisely the coherence order of the system, that is, the difference in Hamming weight between two states). Since a cat-state corresponds to an n-spin, n-quantum coherence state, we may use well established techniques \cite{MQC,MQC2} to create the n-quantum coherence operator, which rotates the fully polarized state $|00 \ldots 0\rangle$ into the cat-state.

In a large magnetic field along the z axis, the dipolar Hamiltonian coupling the spins in a solid takes the form:
\begin{equation}
	\label{DipHam}
 	H_{dip}=\sum_{ij}b_{ij}[\sigma_z^i\sigma_z^j-\frac{1}{2}(\sigma_x^i\sigma_x^j + \sigma_y^i\sigma_y^j)]
 \end{equation}
Under a $\pi/2$ rotation about the y axis, $R_{\frac{\pi}{2}|_{y}}$, the dipolar Hamiltonian is transformed to:
\begin{equation}
	\label{DipHamx}
 \begin{array}{l}
	R_{\frac{\pi}{2}|_{y}}(H_{dip})=\sum_{ij}b_{ij}\frac{3}{8}(\sigma_+^i\sigma_+^j + \sigma_-^i\sigma_-^j)-\frac{1}{2}H_{dip}
 \end{array}
\end{equation}
With an appropriate sequence of rf pulses and delays we can isolate the first term in Eq. \ref{DipHamx}. This operator is an example of Grade Raising (GR) Hamiltonians, operators that increase the coherence order of a state:
\begin{equation}
	\label{DQHam}
	H_{GR}^{(2)}=\sum_{ij}b_{ij}(\sigma_+^i\sigma_+^j + \sigma_-^i\sigma_-^j)
\end{equation}
Acting with this Hamiltonian on a 2-body operator such as the dipolar Hamiltonian we can create higher order GR operators. 
Phase modulated combinations of $H_{GR}^{(2)}$ and $H_{dip}$ have been already shown \cite{Warren} to permit straightforward synthesis of the pure $n$-body GR Hamiltonian $H_{GR}^{(n)}=\prod_{k=1}^n\sigma^+_k+\prod_{k=1}^n\sigma^-_k$. 
Using the $n$-quantum propagator $U_{NQ}=e^{-i\frac{\pi}{4}H^N_{GR}}$ to create and refocus the cat-state instead of the $\sigma_x$ and \textsc{c-not} gates used in scheme 2, we obtain a lower contrast, C=1 \cite{endnote3}, but requiring only to manipulate the natural Hamiltonian with the available control via rf pulse sequences (Figure 2).

Creating the cat-state is experimentally hard, yet we can envisage even simpler schemes. The form of entanglement that influences the contrast is entanglement of the Amplifier spins with the first Amplifier spin (the one which interacts with the target spin), so that a modification in its state will drive a macroscopic change. Hence, it is not necessary to restrict ourselves to the preparation of a cat-state, as a wider class of entangled states is equally useful. To ensure that the first Amplifier spin is entangled with the rest of the spins, we must operate on the system with a GR operator that always contains this first spin. The GR Hamiltonian used previously as a basic tool to create entanglement is now replaced by:
\begin{equation}\label{GR1}
H_{GR1}^{(2)}=\sum_i b_{1i}(\sigma^+_1\sigma^+_i + \sigma^-_1\sigma^-_i).
\end{equation} 
It is easy to create this operator since the coupling between the target and the first spin makes it distinguishable from the other Amplifier spins 
\cite{endnote4}. 
In order to realize a more robust scheme, we can introduce entanglement only conditionally on the state of the target spin, by applying $H_{GR1}^{(2)}$ conditionally on the target spin state. 

Even with these generalizations, the experimental task is still demanding: We need to apply the inverse of a rather complex map, but this inversion could dramatically amplify small errors done in its implementation. It has been actually shown that the fidelity (or overlap) between an arbitrary initial state evolved under a sufficiently random unitary operator and this same state evolved under a perturbed version of the operator decays exponentially with the number of repetitions of the map \cite{FidDecay}. We can nonetheless turn these considerations to our advantage by coupling the chaotic dynamics to the single spin. If a perturbation is applied conditionally on the state of the target spin, the resultant states will have an exponentially decreasing fidelity. Since the expectation value of the magnetization is a much weaker measurement than the fidelity, we expect the exponential rate to simply bound the contrast growth rate.

Because of the local nature of the spin-spin interactions, the perturbation will be limited to the closest neighbors of the target spin (it could be, for example, the $H_{GR1}^{(2)}$ operator). At a repetition $r$ of the map, the perturbation will have affected only a factor-space of the total Hilbert space of size $N_r$, giving a completely randomized state and an almost zero polarization for the subsystem considered. If the evolution map is sufficiently random (and thus does not have excessive symmetries), the size $N_r$ can grow to a large fraction of the Hilbert space with the increasing number of repetitions, allowing for a good contrast to be observed.

\begin{figure}[b]
	\centering
		\includegraphics[scale=0.31]{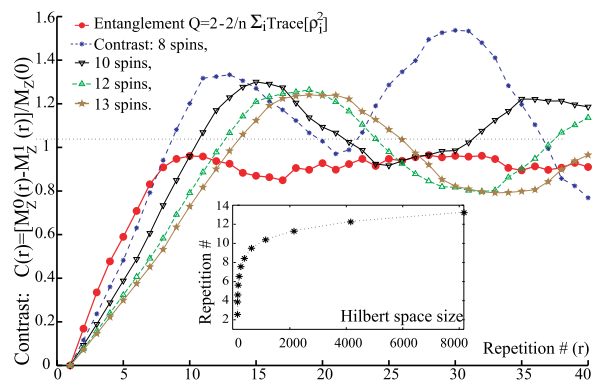}
	\caption{Entanglement (10 spins) and contrast for different number of spins. In the inset: number of repetitions to reach contrast $\approx 1$ as a function of the Hilbert space size, showing a logarithmic dependence. The (perturbed) map applied applied to the initially fully polarized Amplifier is: $U= e^{-itH_{GR1}^{(2)}}e^{-iTH_{dip}}$, with: $tb_{1,2}=1/2$ and $Tb_{1,2}=\pi/\sqrt{2}.$}
	\label{Polarization}
\end{figure}
A pseudo-random map possessing these features can be implemented in a solid state NMR system using the dipolar Hamiltonian (Eq. 1) propagator as the unperturbed map and as perturbation the $H_{GR1}^{(2)}$ operator (Eq. 4), applied conditionally on the state of the target spin. This many body evolution provides a complex enough dynamics, even in the presence of residual symmetries in the Hamiltonian. 

We can explain the effects of this procedure also in terms of creation of entanglement localized around the first neighbor that is then spread out by spin diffusion\cite{Bloembergen}. These two successive steps, repeated many times, ensure that the change affects all the Hilbert space, creating a global entangled state. A good measure of global entanglement is the average loss of purity upon tracing over one spin: $Q=2-\frac{2}{n}\sum_i Tr\{\rho_i^2\}$ \cite{Meyer}, where $\rho_i$ is the partial trace over the $i^{th}$ spin. For a maximally entangled state ($Q=1$), the polarization is zero.

Because of the complexity of the dynamics involved, we found no simple analytic result to predict the contrast growth. We ran instead simulations for a limited number of spins in a linear chain, following the polarization and the entanglement as a function of the number of repetitions of the two basic steps ($H_{GR1}^{(2)}$ and $H_{Dip}$).

From the simulations we can observe the predicted dynamics: Regions of the physical space farther away from the first spin are excited later in time. 
When the entanglement reaches $Q\approx 1$ and the polarization goes to $\approx 0$, we have a saturation phenomenon, and except for fluctuations, that decrease in amplitude with the number of spins, the macroscopic properties of the state no longer change. This effect, caused by the finite size of the Hilbert space, is also observed in the fidelity decay, that for very strong perturbation saturates at $1/N$, with $N$ the size of the Hilbert space.
Notice that we reach contrast $\approx 1$ for a number of repetitions on the order of the spin number, as expected, since at each cycle only one quantum of polarization can be changed by the action of the GR-Hamiltonian in a 1D chain of spins. We expect a faster rate in 2 and 3 dimensions, due to a higher number of first neighbors. 
\begin{figure}[hb]
	\centering
			\includegraphics[scale=.85]{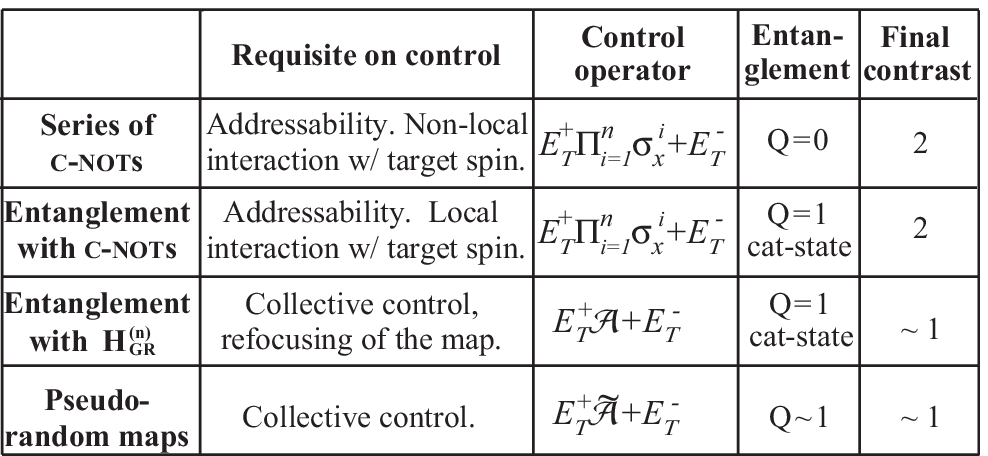}		
\end{figure}

In conclusion, we have shown that it is possible to transfer polarization from a target spin to an ensemble of spins, through the creation of a highly entangled state with rf pulses and the coupling between the target spin and its closest neighbors. 
The characteristics of the different methods discussed range from optimal contrast but with extreme control on the system to a lower contrast and more accessible experimental conditions (as summarized in the table). In particular, the last two methods can find an immediate application to the signal enhancement of rare spins, embedded in a sample containing a more abundant spin species: these schemes require only collective control on the ensemble spins and the number of operations needed is moderate if the ratio of rare to abundant spins is large enough. 
The methods and physical systems proposed open the possibility to a new class of devices, where quantum effects, such as entanglement, are used to make a transition from microscopic to macroscopic properties.
This work was supported by NSF, DARPA and  ARDA/ARO.


\begin{thebibliography}{99}
\bibitem{endnote1} Restricting the measurement to only two orthogonal states is not a limitation in the context of QIP, since it is sufficient for the read-out stage \cite{DiVincenzo}.
\bibitem{endnote2} The "no-cloning" theorem Cloning does not apply here, since the target spin is already in one of two orthogonal states.
\bibitem{endnote3} The total operator $U=U_{GR}^{(n)}U_{cnot}U_{GR}^{(n)\dag}$ $=E_0^-+E_0^+(\sigma_x^1+(\frac{\sqrt 2}{2}-1)\sigma_x^1(\prod_{i=2}^nE^+_i+\prod_{i=2}^nE^-_i)$ $-i\frac{\sqrt 2}{2}\sigma_z^1(\prod_{i=2}^n\sigma^+_i+\prod_{i=2}^n\sigma^-_i))$ either conserves the magnetization if $|\psi\rangle=|0\rangle$ or brings it to zero if $|\psi\rangle=|1\rangle$.
\bibitem{endnote4}If we repeat the pulse sequence for the $H_{GR}^{(2)}$ with a $\pi/2$ phase shift, the overall evolution would be completely refocused; by inverting the closest spin phase in between the two sequences, only the terms containing the first spin will survive, creating the GR1 Hamiltonian: $e^{-iH_{GR}^{(2)}t}e^{-i\frac{\pi}{2}\sigma_z^1}e^{iH_{GR}^{(2)}t}\approx e^{-2iH_{GR1}^{(2)}t}$.
\end{thebibliography}
\end{document}